\documentclass[twocolumn]{mn2e}
\usepackage{graphicx}
\title{Testing MOND with Ultra-Compact Dwarf Galaxies}

\author[Riccardo Scarpa]{Riccardo Scarpa\\
       European Southern Observatory, 3107 Alonso de Cordova, 
       Santiago, Chile}

\date{}

\begin{document}

\maketitle

\begin{abstract}
The properties of the recently discovered Ultra-Compact
Dwarf Galaxies (UCDs) show that their internal acceleration of gravity
is everywhere above $a_0$, the MOdified Newtonian Dynamics (MOND)
constant of gravity.  MOND therefore makes the strong prediction that
no mass discrepancy should be observed for this class of objects. This
is confirmed by the few UCDs for which virial masses were derived.  We
argue that UCD galaxies represent a suitable test-bench for
the theory, in the sense that even a single UCD galaxy showing
evidence for dark matter would seriously question the validity of
MOND.
\end{abstract}

\section{Introduction}

Dwarf galaxies have attracted the attention of astrophysicists in the
last years for many reasons. Among others, these galaxies are found to
contain an impressive amount of non-baryonic dark matter having
mass-to-light ratios up to 100 and being dark matter dominated all the
way to their center (Mateo 1988, Kleyna et al. 2001, 2004).

The discovery (Hilker et al. 1999; Drinkwater et al. 2000; Phillips et
al. 2001) in an all-object survey of the Fornax cluster of a new type
of galaxies, the ultra-compact dwarfs (UCD), have added a new member
to the dwarf family.  Unlike other dwarfs, however, UCD galaxies have
very high central concentration and thus star-like morphology in
typical one-arcsecond-resolution ground-based imaging. At the distance
of the Fornax cluster (20 Mpc; Hubble constant H$_0$ = 75 km s-1
Mpc-1) this implies sizes of 100 pc at most.  The absolute magnitudes
range from $-11<$M$_B<-13$, significantly brighter than the
brightest globular cluster known (Meylan et al. 2001).  Thus, UCDs are
squarely placed midway between regular dwarf galaxies and globular
clusters. 

Recent determination (Drinkwater et al. 2003) of their
effective radius (ranging from 10 to 22 pc) and velocity dispersion
(ranging from 24 to 37 km s$^{-1}$), indicates that, if any, there is
very little dark matter in UCD galaxies.  Indeed assuming UCDs are
virialized structure, masses ranging from $10^7$ to $10^8$ M$_\odot$
are found (Drinkwater et al. 2003). The corresponding mass-to-light
ratio varies from 2 to 4 in solar units, fully consistent with the
expectation for an old stellar population without significant amount
of dark matter.

In the literature two different scenarios have been proposed to
explain why UCDs show no mass discrepancy. One possibility is that
UCDs are giant globular clusters (Mieske et al. 2002), possibly the
result of the merging of the giant stellar clusters created during
periods of strong galaxy interaction (Fellhauer and Kroupa 2002), like
the one observed in the ``Antennae'' system (e.g., Whitmore et
al. 1999).  The other possibility is suggested by the fact that UCDs
luminosity and size match well the one of the {\it nucleus} of dwarf
galaxies. It is therefore possible that UCD galaxies are the remnant
of normal nucleated dwarf galaxies that have lost their external halo
and dark matter,
because of strong and repeated interaction with other galaxies. This
latter scenario is referred to as galaxy threshing (Gregg et al. 2003).

To these two possibilities to explain the low M/L ratio of UCDs, 
I would add a third one: Modified Newtonian
Dynamics (MOND; Milgrom 1983).  MOND posits the breakdown of
Newtonian dynamics when the acceleration of gravity goes below $a_0 =
1.2\times 10^{-8}$ cm s$^{-2}$.  A hypothesis based on the fact that
mass discrepancies are observed in stellar systems when and only when
the internal acceleration of gravity falls below $a_0$ (e.g., Binney
2004).  Despite many
attempts, MOND resisted stubbornly to be falsified as an alternative
to cold dark matter (CDM; e.g. Sanders, McGaugh 2002) and succeeds in
explaining the properties of an increasingly large number of stellar
systems without invoking the presence of non-baryonic dark matter.
Now that a Lorentz covariant theory for MOND exists (Bekenstein 2004),
there is an obvious need to search for more sites where predictions of
MOND and CDM could clash. Ultra-Compact Dwarf galaxies may represent
one of them.

\section{MOND prediction for UCD galaxies}

Considering the most extreme case of M$_B =-11$, a radius of 100 pc,
and M/L=2 in solar units, we see that even at the outer edge the
acceleration of gravity is $5 \times 10^{-8}$ cm s$^{-2} >
a_0$. According to MOND, then, UCDs are everywhere in Newtonian
regime. The simple and at the same time
strong prediction is that {\it no non-baryonic dark matter is to be
found in UCDs}.
Up to now, MOND is in full agreement with observations.  

Within this
framework, the reason why we do not observe a mass discrepancy is
neither due to the particular evolutionary history of this galaxies,
nor to the fact that UCDs can actually be giant globular clusters. It
is solely due to the fact that they are in Newtonian regime and
therefore should strictly obey Newtonian dynamics.

Based on this simple consideration, one has to conclude that UCDs
provide a very good test-bench for MOND. It is sufficient to find a
single UCD with an unquestionably large mass discrepancy to falsify
the theory.  Targeting stellar-like objects in the right range of
luminosity with multi-object spectrographs will certainly allow in the
near future the discovery of many more of this galaxies.  Indeed, a
number of UCD in Virgo has been already found (Drinkwater et
al. 2004), though their velocity dispersion is still unknown.  Once a
seizable sample will be available, beside the interest of better
investigate the properties of UCDs, we will be able to put strong
constraints on MOND.

Moreover, in the case some intermediate UCD galaxies will be found,
i.e., objects with a small halo thus resembling a little more regular
nucleated dwarf, MOND made another interesting prediction.  If it will
possible to trace their velocity dispersion profile sufficiently far
away from the center, then a mass discrepancy will be observed as soon
as internal acceleration of gravity below $a_0$ are probed.  A similar
behavior has been already observed in three galactic globular clusters
(Scarpa, Marconi, and Gilmozzi 2003, 2004).

%\begin{acknowledgments}
I wish to thanks F. Patat, G. Marconi, and M. Romaniello for helpful
discussion and suggestion.
%\end{acknowledgments}

\end{document}